\documentclass[twoside,onecollarge]{article}
 
\ifx\pdfoutput\undefined
   \usepackage[dvips]{graphicx}
  \else%
   \usepackage[pdftex]{graphicx}%
   \fi
   \usepackage{epstopdf}

\usepackage[applemac]{inputenc}
\usepackage[english]{babel}
\usepackage{amsfonts}
\usepackage[reqno]{amsmath}
\usepackage{float}
\usepackage{bbm}
\usepackage{color}

\makeindex

\newtheorem{proposition}{Proposition}

\newcommand{\R}{\mathbb{R}}

\newcommand{\E}{\mathbb{E}}
\newcommand{\V}{\mathbb{V}}

\newcommand{\rnf}{\renewcommand{\thefootnote}{\arabic{footnote}}}
\newcommand{\thankyou}[2]{\stepcounter{footnote}\footnotetext[#1]{#2}}

\title{Optimal starting times, stopping times and risk measures for algorithmic trading: Target Close and Implementation Shortfall}
\author{\rnf Mauricio Labadie
              \footnotemark[1]\quad
              \footnotemark[3]
\and
\rnf Charles-Albert Lehalle
              \footnotemark[2]\quad
              \footnotemark[3]
}

\date{\today}

\begin{document}
\DeclareGraphicsExtensions{.pdf,.gif,.jpg} \maketitle

\thankyou{1}{\,Corresponding author. EXQIM (EXclusive Quantitative Investment Management). 24 Rue de Caumartin 75009 Paris, France. Email: mauricio.labadie@gmail.com.}

\thankyou{2}{\,CFM (Capital Fund Management). 23 Rue de l'Universit\'e 75007 Paris, France.}

\thankyou{3}{\, This research was mostly done when both authors were working at Cr\'edit Agricole - Cheuvreux (now Kepler Cheuvreux).}



\maketitle

\begin{abstract}
We derive explicit recursive formulas for Target Close (TC) and Implementation Shortfall (IS) in the Almgren-Chriss framework. We explain how to compute the optimal starting and stopping times for IS and TC, respectively, given a minimum trading size. We also show how to add a minimum participation rate constraint (Percentage of Volume, PVol) for both TC and IS.\\

We also study an alternative set of risk measures for the optimisation of algorithmic trading curves. We assume a self-similar process (e.g. L\'evy process, fractional Brownian motion or fractal process) and define a new risk measure, the $p$-variation, which reduces to the variance if the process is a Brownian motion. We deduce the explicit formula for the TC and IS algorithms under a self-similar process.\\

We show that there is an two-way relationship between self-similar models and a family of risk measures called $p$-variations. indeed, it is equivalent to have (1) a self-similar process and calibrate empirically the parameter $p$ for the $p$-variation, or (2) a Brownian motion and use the $p$-variation as risk measure instead of the variance. We also show that $p$ can be seen as a fine-tuning parameter which modulates the aggressiveness of the trading protocole: $p$ increases if and only if the TC algorithm starts later and executes faster.\\

Finally, we show how the parameter $p$ of the $p$-variation can be implied from the optimal starting time of TC. Under this framework $p$ can be viewed as a measure of the joint impact of market impact (i.e. liquidity) and volatility.

\end{abstract}

\bigskip


\maketitle

\newpage

\tableofcontents

\newpage

\section{Introduction}

\paragraph{Purpose of the paper.}

Trading algorithms are used by asset managers to control the trading rate of a large order, performing a balance between trading fast to minimise exposure to market risk and trading slow to minimise market impact (for an overview of quantitative trading methods, see \cite{citeulike:10363469} and \cite{citeulike:10363473}). This balance is usually captured via a cost function which takes into account two joint effects, namely the market impact and the market risk. The first frameworks to be proposed were \cite{BLA98} and \cite{OPTEXECAC00}, the latter using a mean-variance criteria. More sophisticated cost functions have been already proposed in the academic literature, leading to the use of different optimization approaches like stochastic control (see \cite{citeulike:5797837} or \cite{citeulike:9272221}) or stochastic algorithms (see \cite{citeulike:5177512}).\

From the practitioners' viewpoint, the cost function to choose is far from obvious. The easiest way to proceed is to replace the choice of the cost function by observable features of the market. This is the approach chosen in this paper, where the cost function generalises the mean-variance frameworks of both Almgren-Chriss and \cite{citeulike:6699563}, which a mean-$p$-variation. Instead of complex cost functions and compute-intensive parameter calibration, this paper proposes a simpler approach that covers a large class of parametrized cost function already in use by practitioners. We calibrate the parameters from observable variables like stopping times and maximum participation rates.\

This approach is very flexible and customisable. Indeed, since it depends on a single fine-tuning parameter $p$, a practitioner can either calibrate $p$ or modify it by hand to fit their risk budget. A good example is the maximum participation rate: actually most practitioners are using a mean-variance criteria with an arbitrary risk aversion parameter, but add a ``control layer'' to their algorithms in order to ensure that the participation on real time will never be more than a pre-determined threshold (i.e. the trading algorithm will never buy or sell more than a certain percentage of the volume traded by the whole market); here we propose a way to include this constraint into the full optimisation process, at the very first step of the process. Moreover, some traders know that they would like to see a given algorithm finish a buy of a given number of shares within a certain time period; again; we propose a way to implicit the parameters of the cost function to achieve this.

\paragraph{An optimal trading framework for the target close and implementation shortfall benchmarks with percentage of volume constraints.}\

A TC (Target Close) algorithm is a trading strategy that aims to execute a certain amount of shares as near as possible to the closing auction price. Since the benchmark with respect to which the TC algorithm is measured is the closing price, the trader has interest in executing most of their order at the close auction. However, if the number of shares to trade is too large, the order cannot be totally executed at the close auction without moving the price too much due to its market impact \cite{citeulike:10363463}. Therefore, the trader has to trade some shares during the continuous auction phase (i.e. before the close) following one of the now well-known optimal trading algorithms available, e.g. mean-variance optimisation (following \cite{OPTEXECAC00}) or stochastic control (like in \cite{citeulike:5797837}).\

As we have mentioned above, this paper will stay close to the original Almgren-Chriss framework, extending the risk measure from the variance to a general $p$-variation criterion. The goal of this paper to explain the practical interpretation of the $p$-variation parameter  used in the optimisation scheme and show how to choose them optimally in practice.\

The $p$-variation is an extension of the variance, depending on $p$, since when $p=2$ we recover the variance. When $p\neq 2$ the trader assumes that (1) the price is no longer a martingale, i.e. there are patterns in prices (trend-following or mean-reverting), and (2) the time-scaling properties of prices are not as in the Brownian motion. Therefore, a new risk measure other than variance is needed. This paper explores the impact of $p$ on the properties of the obtained optimal trading curve, and relates it with self-similar processes (e.g. fractional Brownian motion, L\'evy processes and multifractal processes).

\paragraph{Inverting the optimal liquidation problem putting the emphasis on observables of the obtained trading process.}\

We will show that the TC (Target Close) algorithm can be seen as a ``\emph{reverse IS}'' (Implementation Shortfall) --see equation (\ref{eq:f:1}) and following for details--. In this framework, the starting time for a TC is as important than the ending time for an IS. For practitioners this distinction is even more critical since shortening the trading duration of an IS because of an interesting price opportunity can always be justified, but beginning sooner or later than an ``\emph{expected optimal start time}'' for a TC is more difficult to explain.\

The paper also shows that the results obtained for the TC criterion can be applied to the IS criterion because TC and IS are both sides of the same coin. Indeed, on the one hand, the TC has a pre-determined end time, its benchmark is the price at the end of the execution and the starting time is unknown. On the other hand, IS has a pre-determined starting time, its benchmark is the price at the beginning of the execution and the stopping time is unknown. Therefore, there is no surprise that the recursive formula for IS turns out to be exactly the same that for TC but with the time running backwards.\

It is customary for practitioners to put constraints on the maximum participation rate of a trading algorithms (say 20\% of the volume traded by the market). Therefore, it is of paramount importance to find a systematic way of computing the starting time of a TC under a percentage of volume (PVol) constraint. 
Such an ``\emph{optimal trading policy under PVol constraint}'' is properly defined and solved in this paper. A numerical example with real data is provided, where the optimal trading curves and their corresponding optimal starting times are computed.\

Solving the TC problem under constraints allows us to analyze the impact of the parameters of the optimisation criterion on observable variables of the trading process. It should be straightforward for quantitative traders the task to implement our results numerically, i.e. to choose the characteristics of the trading process they would like to target and then infer the proper value of the parameters of the criterion they need.

\paragraph{Link between a mean $p$-variation criterion and self similar price formation processes.}\

\cite{OPTEXECAC00} developed a mean-variance framework to trade IS (Implementation Shortfall) portfolios driven by a Brownian motion. More recently,  \cite{citeulike:5094012} extended the model to Gaussian portfolios whilst \cite{citeulike:10363463} addressed the same problem for the geometric Brownian motion. In this article we extend the analysis to a broad class of non-Brownian models, the so-called self-similar models, which include L\'evy processes and fractional Brownian motion (for empirical studies about the self-similarity of intraday data, see \cite{citeulike:10665572}, \cite{citeulike:8524050} or \cite{citeulike:10665597}). We study in detail the relationship between the exponent of self-similarity, the choice of the risk measure and the level of aggressiveness of the algorithm. We show that there are two opposite approaches that nevertheless give the same recursive trading formula: one assumes a self-similar process, estimates the exponent of self-similarity $H$ and chooses the $p$-variation via $p=1/H$; the other assumes a classical Brownian motion and chooses the $p$-variation as the risk measure instead of the the variance.

In the same way the starting time of a TC or the ending time of an IS can be used as an observable to infer values of parameters of the optimization program, the maximum participation rate is expressed as a function of $p$ for a mean $p$-variance criterion. In the light of this, a quantitative trader who has chosen to trade no more than 30\% of the market volume during a given time interval, can modifie the value of $p$ to fine-tune their execution and respect their constraints.
\\

This paper formalizes an innovative approach of optimal trading based on observable variables, risk budget and participation constraints. By doing so, it opens the door to a framework close to risk neutral valuation of derivative products in optimal trading: instead of choosing the measure under which to compute the expectation of the payoff (because optimal trading is always considered under historical measure), we propose to infer the value of some parameters of the cost function so that the trading process will satisfy some observable characteristics (start time, end time, maximum participation rate, etc). In this framework, instead of being hedged with respect to market prices, the trader will be hedged with respect to the risk-performance profile of an ideal trading process, i.e. a proxy she defined a priori.\

Notice that we have chosen to extend the usual mean-variance criterion rather than going to more non-parametric approaches like stochastic control. The main reason for this approach is because our framework allows more explicit recurrent formulas, not to mention that our method can be easily extended to other execution algorithms besides TC and IS.

\paragraph{Organisation of the article.}\

In Section 2 we derive a nonlinear, explicit recursive formula for both the TC and IS algorithms with a nonlinear market impact. We explain how to build a TC algorithm under a maximum participation rate constraint (percentage of volume, PVol). We provide a numerical example using real data, in which we computed the trading curves and their optimal starting time. All our computations can be also applied to IS.\

In Section 3 we extend the analysis for a class of non-Brownian models called self-similar processes, which include L\'evy Processes, fractional Brownian motion and fractal processes. We define an \emph{ad hoc} risk measure, denoted $p$-variation, which renders the cost functional linear in time. We show numerically that the exponent of self-similarity $H$ can be viewed as a fine-tuning parameter for the level of aggressiveness of the TC algorithm under PVol constraint.\

In Section 4 we assess the effect of the parameter $p$ in terms of risk management. We show the existence of an equivalence between risk measures of $p$-variation type and self-similar models of exponent $H$: choosing a self-similar model, estimating $H$ and defining $p=1/H$ for the risk measure yields the same trading curve as assuming a Brownian motion but changing the risk measure from variance to $p$-variation. We also study the effect of $p$ on the starting times for TC and the slopes of the corresponding trading curves\

We conclude by showing how the parameter $p$ of the $p$-variation can be implied from the optimal starting time of TC. In that framework $p$ can be viewed measure of the joint impact of market impact (i.e. liquidity) and volatility.

\section{Optimal starting and stopping times}

\subsection{A review of the mean-variance optimisation of Almgren-Chriss}\label{sec:ac}

This section recalls the framework, notation and results in \cite{OPTEXECAC00} and \cite{citeulike:5094012}. Suppose we want to trade an asset $S$ throughout a time horizon $T>0$. Assume that we have already set the trading schedule, i.e. we will do $N$ trades at evenly distributed times
\[
0=t_0<t_1<t_2<\cdots<t_N=T.
\]
The goal is, given a volume to execute $v^\ast$, find the optimal quantity of shares $v_n$ to execute at time $t_n$ that minimise the joint effect of market impact and market risk under the constraint
\begin{equation}\label{volume-exec}
\sum_{i=1}^N v_i = v^\ast>0 \,.
\end{equation}
Define $\tau=t_n-t_{n-1}$ and assume that the price dynamics follows a Brownian motion, i.e.
\begin{equation}\label{single-gaussian}
S_{n+1}=S_n+\sigma_{n+1}\tau^{1/2}\varepsilon_{n+1},
\end{equation}
where $(\varepsilon_n)_{1\le n\le N}$ are i.i.d. normal random variables of mean zero and variance $1$, and $(\sigma_n)_{1\le n\le N}$ are the historical volatilities at the trading times $(t_n)_{1\le n\le N}$.\footnote{\, We will be assuming in this paper that the historical volatility and volume curves $(\sigma_n)_{1\le n\le N}$ and $(V_n)_{1\le n\le N}$ are non-costant throughout the day and known ex-ante e.g. as the average over a period of time. In practice, both the average historical volatility and volume present a U-shape pattern: they are higher at the open and close of the market than in between.} Following \cite{OPTEXECAC00} and \cite{citeulike:5094012}, we will model the temporary market impact as a function $h(v_n)$, i.e. depending solely on what happens at each trading time.\footnote{\, Strictly speaking, the function $h(\cdot)$ is the temporary market impact. In our study we have decided to neglect the permanent market impact for two main reasons. First, the permanent impact is smaller than the temporary impact since we are not taking into account the relaxation due to the elasticity of prices. Therefore, we are overvaluing the real impact in the long run, and as such our analysis is thus conservative. Second, since the permanent impact is usually modelled as linear, any trading curve will have the same permanent impact. There is thus no loss of generality if we assume zero permanent impact.} Under this framework, the wealth process (i.e. the full trading revenue upon completion of all trades) is
\begin{eqnarray}\label{single-wealth}
W &=& \sum_{n=1}^N q_n v_n(S_n+q_n h(v_n))\\
 &=& \sum_{n=1}^N q_n v_n S_n+\sum_{n=1}^N v_n h(v_n)\,,\nonumber\\
 &=& S_0\sum_{n=1}^N q_n v_n + \sum_{n=1}^N\sum_{i=1}^n q_n v_n \sigma_i\tau\varepsilon_i+\sum_{n=1}^N v_n h(v_n)\,,\nonumber
\end{eqnarray}
where $q_n=1$ if we buy at time $t_n$ and $q_n=-1$ if we sell. Assume we have long-only portfolio, i.e. $q_n=+1$. In this framework, if we use the identity
\[
\sum_{n=1}^N\sum_{i=1}^n a_nb_i = \sum_{n=1}^N b_n\left(\sum_{i=n}^N a_i\right)
\]
and the change of variables
\[
x_n:=\sum_{i=n}^N v_i\quad \iff\quad v_n = x_n - x_{n+1}
\]
it follows that the wealth process becomes
\begin{equation}\label{single-wealth-long}
W(x_1,\dots,x_n)=S_0v^\ast + \sum_{n=1}^N\sigma_n\tau^{1/2}\varepsilon_n x_n + 
\sum_{n=1}^N(x_n-x_{n+1})h(x_n-x_{n+1}).
\end{equation}
The expectation and variance of the wealth process \eqref{single-wealth} are, respectively,
\begin{equation}\label{single-exp-var}
\E(W) = S_0v^\ast+\sum_{n=1}^N(x_n-x_{n+1})h(x_n-x_{n+1})\,,\qquad \V(W) = \sum_{n=1}^N\sigma_n^2\tau x_n^2\,.\nonumber
\end{equation}
Therefore, the corresponding mean-variance cost functional for a level of risk aversion $\lambda>0$ is
\begin{eqnarray}\label{single-functional}
J_\lambda(x_1,\dots,x_N) &=& \E(W)+ \lambda\V(W)\\
 &=& S_0v^\ast +\sum_{n=1}^N(x_n-x_{n+1})h(x_n-x_{n+1})+ \lambda\sum_{n=1}^N\sigma_n^2\tau x_n^2\,.\nonumber
\end{eqnarray}
In order to find the optimal trading curve we have to find the points $(x_1,\dots,x_N)$ that solve the system \[
\frac{\partial J_\lambda}{\partial x_n}=0\,,\quad n=1,\dots,N\,.
\]
If the market impact function $v_n\mapsto h(v_n)$ is strictly monotone and differentiable for positive values, e.g. $h(s)=s^\gamma$ with $\gamma>0$, it is possible to obtain explicit recursive algorithms of the form
\begin{equation}\label{eq:euler}
x_{n+1} = f(x_n,x_{n-1})
\end{equation}
constraints $x_0=v^\ast$ and $x_{N+1}=0$.

\subsection{The Shooting Method}

Notice that \eqref{eq:euler} is completely determined once the values $x_0$ and $x_1$ are known. More generally, once two different values $x_i$ and $x_j$ with $i<j$ are known, the other values $x_n$ can be computed. However, the method is not necessarily explicit and recursive if $i\neq 0$ or $j\neq 1$. In our case, we have the constraints $x_0=v^\ast$ and $x_{N+1}=0$, i.e. an initial and a final condition, which implies that \eqref{eq:euler} is no longer an explicit and recursive algorithm. Nevertheless, the problem can be solved explicitly and recursively using a dichotomy method called \emph{Shooting Method}.\\

We start with the following ordinary Differential Equation (ODE):
\begin{equation}\label{initial-value}
y^{\prime\prime}=g(y,y^\prime); \quad t\in[a,b],\quad y(a)=A,\quad y^\prime(a)=\alpha,
\end{equation}
where $g$ is a bounded and differentiable function. According to the standard theory of ODEs, the initial-value problem \eqref{initial-value} has a unique solution $y(t)$.\footnote{\,Strictly speaking, the solution $y(t)$ only exists locally, but it is globally defined if $g$ and all its partial derivatives are continuous and bounded (see e.g. \cite{book:perko}).}\

Now consider the boundary problem
\begin{equation}\label{boundary}
y^{\prime\prime}=g(y,y^\prime); \quad t\in[a,b],\quad y(a)=A\in\R,\quad y(b)=B\in\R.
\end{equation}
It is not evident that \eqref{boundary} has a solution. However, we can try to translate the boundary problem \eqref{boundary} into an initial-value problem of type \eqref{initial-value}, for which we know that solutions do exist.\

\begin{figure}[htbp]
\begin{center}
\includegraphics[height=2.5in]{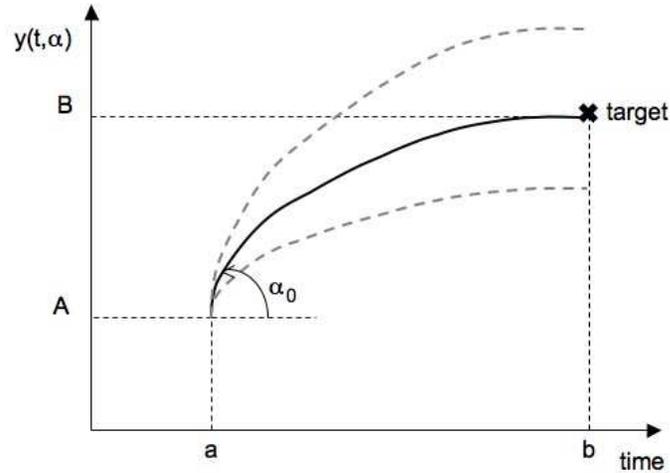} 
\caption{\small{The shooting method. Varying the ``shooting angle'' $\alpha$ we can find the right angle $\alpha_0$ such that the curve $y(t,\alpha)$ ''hits the target'', i.e. $y(b,\alpha_0)=B$.}}
\label{fig-shooting}
\end{center}
\end{figure}

The shooting method consists exactly in this translation. Indeed, for any $\alpha\in\R$, the initial-value problem \eqref{initial-value} has a solution $y(t,\alpha)$. To solve the boundary problem \eqref{boundary}, we need to find $\alpha_0$ such that $y(b,\alpha_0)=B$. Roughly speaking, we are playing with the parameter $\alpha$ in order to ``hit'' $B$ (see Figure \ref{fig-shooting}). In consequence, the boundary problem \eqref{boundary} reduces to find a zero of the function
\[
F(\alpha)=y(b,\alpha)-B,
\]
which can be solved using any numerical method, e.g. bisection or Newton (see e.g. \cite{book:stoer}).\

Suppose we have already found the optimal trading curve $(x_1,\dots,x_N)$ via a recursive algorithm of the form, i.e.
\begin{equation}\label{rec-algo}
x_{n+1}=F(x_n,x_{n-1})
\end{equation}
under the constraints $x_0=1$ and $x_{N+1}=0$. Using an induction argument it can be shown that $x_n$ is a function of $x_0$ and $x_1$ for $n\ge2$. By induction we obtain that $x_{N+1}$ is a function of $x_1$, i.e.
\[
x_{N+1}=F(x_1),
\]
because $x_0$ has been already fixed to be equal to $1$. If we define $x_1=\alpha$  then $\alpha$ is a free parameter that completely determines the optimal trading curve. An important remark is that $\alpha$ can be related to the the slope of the trading curve at $x_0$ because
\[
\frac{x_1-x_0}{1-0}=\alpha-1.
\]
By considering $\alpha$ as the ``slope'' we can see an analogy of the optimal trading curve with the shooting method. Under this new framework, our optimization problem reduces to find $\alpha_0$ such that $F(\alpha_0)=0(=x_{N+1})$. Indeed, if we choose $x_1 = \alpha_0$ then $x_{n+1}=F(\alpha_0)=0$, as we wished.\

The beauty of the analogy with the shooting method is that we are working with a 1-dimensional function $F(x_1)$ instead of the $N$-dimensional functional $J_\lambda(x_1,\dots,x_N)$. In consequence, using the shooting method we are always solving a 1D problem regardless of the number of trades $N$. This fact renders our algorithm very appealing for high frequency trading.\


\subsection{Derivation of the Target Close (TC) algorithm}\label{sec:tc}


As in \cite{almgren03},  \cite{almgren03} and \cite{bouchaud10}, we will consider a power market impact function, i.e.
\begin{equation}\label{market-impact}
h(v_n)=\kappa\sigma_n\tau^{1/2}\left(\frac{v_n}{V_n} \right)^\gamma\,,
\end{equation}
where $v_n$ is the amount of shares executed at the $n$-th pillar (i.e. at time $t_n$), $V_n$ is the historical volume at the $n$-th pillar, $\sigma_n$ is the (normalised) historical volatility at the $n$-th pillar, and $\kappa$ and $\gamma$ are positive constants. Under this framework, the wealth process \eqref{single-wealth-long} takes the form
\begin{equation}\label{eq-wealth}
W = S_0v^\ast + \sum_{n=1}^N\sigma_n\tau^{1/2}\varepsilon_n x_n + \sum_{n=1}^N \kappa \sigma_n\tau^{1/2} v_n \left(\frac{v_n}{V_n}\right)^\gamma \,,
\end{equation}
where $N$ is the number of slices in the trading algorithm. The first term in the right-hand side of \eqref{eq-wealth} is the cost of executing $v^\ast:=\sum_{i=1}^N v_i$ shares; the second term models the market impact of the execution as a power law of the percentage of volume executed at each pillar $n=1,\dots,N$.\\

For a TC algorithm, the benchmark is the closing price. Therefore, the wealth process relative to this benchmark is
\begin{eqnarray*}
W^\sharp &=& W - S_N\sum_{n=1}^N v_n\,.
\end{eqnarray*}
Define
\[
x_n := \sum_{i=1}^n v_i \quad\textnormal{and}\quad y_n := \sum_{i=n}^N v_i
\]
(notice that this notation differs from what we used in Section \ref{sec:ac}). Under this framework, it follows that
\begin{eqnarray}\label{eq:f:1}
 W^\sharp &=& W - S_0v^\ast - y_1\sum_{n=1}^N\sigma_n\tau^{1/2}\varepsilon_n \nonumber\\
 &=& \sum_{n=1}^{N} (y_n-y_1)\sigma_n\tau^{1/2}\varepsilon_n+\sum_{n=1}^N \kappa \sigma_n\tau^{1/2} \frac{(y_n-y_{n+1})^{\gamma+1}}{V_n^\gamma}\nonumber\\
 &=& \left( -\sum_{n=1}^{N-1} x_n\sigma_{n+1}\varepsilon_{n+1}+\sum_{n=1}^N \kappa \sigma_n \frac{(x_n-x_{n-1})^{\gamma+1}}{V_n^\gamma}\right)\tau^{1/2}\,.
\end{eqnarray}
Since the time-step $\tau^{1/2}$ is a constant multiplicative factor, we can consider a normalised relative wealth
\[
\tilde W := \frac{W^\sharp}{\tau^{1/2}}\,.
\]
We are not losing any generality with the normalisation because it is equivalent to use a normalised volatility $\tilde\sigma_n:=\sigma_n\tau^{1/2}$. Under this new framework, the average and variance of $\tilde W$ are, respectively,
\[
\mathbb{E}(\tilde W)=\sum_{n=1}^N \kappa\sigma_n \frac{(x_n-x_{n-1})^{\gamma+1}}{V_n^\gamma}\,,\qquad\mathbb{V}(\tilde W)=\sum_{n=1}^{N-1} x_n^2\sigma_{n+1}^2\,.
\]
The corresponding mean-variance functional is thus
\begin{eqnarray*}
J_\lambda(x_1,\dots,x_N) &=& \mathbb{E}(\tilde W)+\lambda\mathbb{V}(\tilde W)\\
 &=& \sum_{n=1}^N \kappa\sigma_n \frac{(x_n-x_{n-1})^{\gamma+1}}{V_n^\gamma}+\lambda\sum_{n=1}^{N-1} x_n^2\sigma_{n+1}^2\\
 &=& \sum_{n=1}^N \kappa\sigma_n \frac{(x_n-x_{n-1})^{\gamma+1}}{V_n^\gamma}+\sum_{n=1}^N \lambda(n,N) x_n^2\sigma_{n+1}^2\,,
\end{eqnarray*}
where $\lambda(n,N)=\lambda$ if $n<N$ and $\lambda(N,N)=0$. The optimal trading curve is determined by solving
\[
\frac{\partial J_\lambda}{\partial x_n}=0\,,\qquad n=1,\dots,N\,,
\]
i.e.
\[
\kappa\sigma_n(\gamma+1)\frac{(x_n-x_{n-1})^\gamma}{V_n^\gamma}
-\kappa\sigma_{n+1}(\gamma+1)\frac{(x_{n+1}-x_n)^\gamma}{V_{n+1}^\gamma}+2\lambda(n,N)\sigma_{n+1}^2x_n=0\,.
\]
Returning to the variables $v_n$ we get
\[
\kappa\sigma_n(\gamma+1)\left(\frac{v_n}{V_n}\right)^\gamma
-\kappa\sigma_{n+1}(\gamma+1)\left(\frac{v_{n+1}}{V_{n+1}}\right)^\gamma
+2\lambda(n,N)\sigma_{n+1}^2\left(\sum_{i=1}^n v_i\right)=0\,.
\]
Finally, we obtain an explicit, nonlinear recursive formula of the optimal trading curve for a TC algorithm:
\begin{equation}\label{tc-algo-1}
v_{n+1}=V_{n+1}\left[\frac{\sigma_n}{\sigma_{n+1}}\left(\frac{v_n}{V_n}\right)^\gamma+
\frac{2\lambda(n,N)}{\kappa(\gamma+1)}\sigma_{n+1}\left(\sum_{i=1}^n v_i\right)\right]^{1/\gamma}\,.
\end{equation}

\subsection{Derivation of the Implementation Shortfall (IS) algorithm}

For an IS algorithm, the starting time is given and we have to find the optimal stopping time for our execution. Since the benchmark is the price at the moment when the execution starts, the relative wealth of an IS algorithm is\footnote{\, $S_0$ is the price at time $t_0$ i.e. the moment when the trader decided to start their execution. $S_1$ is the price at $t_1$ i.e. the moment the first trade took place. If we take into account the price slippage due to the delay between $t_0$ and $t_1$ then the benchmark for IS should be $S_0$. On the contrary, if we neglect the delay then the benchmark is $S_1$. In this paper we take the second approach.}
\[
W^\sharp = W-S_1\sum_{n=1}^N v_n\,.
\]
Using the change of variables
\[
x_n:=\sum_{i=n}^N v_i\quad \iff\quad v_n = x_n - x_{n+1}\,,
\]
and equation \eqref{eq-wealth} it can be shown that the relative wealth process is
\begin{eqnarray*}
W^\sharp &=& \sum_{n=2}^N x_n\sigma_n\tau^{1/2}\varepsilon_n+\sum_{n=1}^N \kappa \sigma_n\tau^{1/2} \frac{(x_n-x_{n-1})^{\gamma+1}}{V_n^\gamma}\\
 &=& \left( \sum_{n=2}^N x_n\sigma_n\varepsilon_n+\sum_{n=1}^N \kappa \sigma_n \frac{(x_n-x_{n-1})^{\gamma+1}}{V_n^\gamma}\right)\tau^{1/2}\,.
\end{eqnarray*}
As in the TC case, we can consider a normalised relative wealth
\[
\tilde W := \frac{W^\sharp}{\tau^{1/2}}\,,
\]
whose mean and variance are, respectively,
\[
\mathbb{E}(\tilde W)=\sum_{n=1}^N \kappa\sigma_n \frac{(x_n-x_{n+1})^{\gamma+1}}{V_n^\gamma}\,,\qquad\mathbb{V}(\tilde W)=\sum_{n=2}^N x_n^2\sigma_n^2\,.
\]
In consequence, the corresponding mean-variance functional is
\begin{eqnarray*}
J_\lambda(x_1,\dots,x_N) &=& \mathbb{E}(\tilde W)+\lambda\mathbb{V}(\tilde W)\\
  &=& \sum_{n=1}^N \kappa\sigma_n \frac{(x_n-x_{n+1})^{\gamma+1}}{V_n^\gamma}+\sum_{n=1}^N \lambda(n,1) x_n^2\sigma_n^2\,,
\end{eqnarray*}
where $\lambda(n,1)=\lambda$ if $n>1$ and $\lambda(1,1)=0$. The optimal trading curve is determined by solving
\[
\frac{\partial J_\lambda}{\partial x_n}=0\,,\qquad n=1,\dots,N\,,
\]
i.e.
\[
\kappa\sigma_n(\gamma+1)\frac{(x_n-x_{n+1})^\gamma}{V_n^\gamma}
-\kappa\sigma_{n-1}(\gamma+1)\frac{(x_{n-1}-x_n)^\gamma}{V_{n+1}^\gamma}+2\lambda(n,1)\sigma_n^2x_n=0\,.
\]
Returning to the variables $v_n$ we get
\[
\kappa\sigma_n(\gamma+1)\left(\frac{v_n}{V_n}\right)^\gamma
-\kappa\sigma_{n-1}(\gamma+1)\left(\frac{v_{n-1}}{V_{n+1}}\right)^\gamma
+2\lambda(n,1)\sigma_n^2\left(\sum_{i=n}^N v_i\right)=0\,.
\]
We thus obtain the recursive nonlinear formula for the optimal IS trading curve:
\begin{equation}\label{tc-algo-is}
v_{n-1}=V_{n-1}\left[\frac{\sigma_n}{\sigma_{n-1}}\left(\frac{v_n}{V_n}\right)^\gamma+
\frac{2\lambda(n,1)}{\kappa(\gamma+1)}\frac{\sigma_n^2}{\sigma_{n-1}}\left(\sum_{i=n}^N v_i\right)\right]^{1/\gamma}\,.
\end{equation}

\subsection{Comparison between TC and IS}

If the volatility is constant then the recursive algorithm \eqref{tc-algo-1} for TC  is exactly the same as \eqref{tc-algo-is} for IS, except for the time in IS \emph{running backwards}. In the light of this \emph{mirror} property, the analysis we will be performing fro TC can be nturally extrapolated for IS, e.g. adding a maximum participation rate constraint and computing the optimal starting time.\

If the volatility is not constant then there is a slight difference in the formulas \eqref{tc-algo-1} and \eqref{tc-algo-is}, namely a factor $\sigma_{n+1} (=\sigma_{n+1}^2/\sigma_{n+1})$ for TC and a factor of $\sigma_n^2/\sigma_{n-1}$. From the practitioner's point of view, this difference is paramount. For TC it is the \emph{forward volatility} $\sigma_{n+1}$ which determines the weight of $\sum_{i=1}^n v_i$ i.e. the shares already executed. This comes from the fact that the ending time is fixed (the market close) and the number of shares to trade per pillar increases in time. In this scenario, not only the trader has little room to change their schedule, but also they have to anticipate the volatility one step ahead in order to avoid nasty surprises. For IS it is the \emph{spot volatility} $\sigma_n$ that counts: since the number of shares per pillar decreases in time, the trader has more room to change the trading schedule the closer they are to the end of the execution. In consecuence, they can capitalise on potential arbitrage opportunities.

\subsection{Adding constraints: Percentage of Volume (PVol)}

The TC algorithm can have a \emph{participation rate} constraint, meaning that the size of each slice cannot exceed a fixed percentage of the available volume (either current or historic average). This restriction is called Percentage of Volume (PVol), which is an execution algorithm itself. Under a PVol constraint, the trading slices $v_n$ of the TC algorithm satisfy the constraint
\[
v_n\le q V_n\,,\qquad q\in(0,1)\,.
\]
It is worth to notice that the PVol algorithm is not a solution of the Almgren-Chriss optimisation. Indeed, if it were then
\[
\frac{v_n}{V_n}=p\quad \forall n=1,\dots,N\,,
\]
and from \eqref{tc-algo-1} we would have that
\[
\sum_{i=1}^n v_i=0\,.
\]
In consequence, since $v_n\ge0$ it follows that $v_n=0$ for all $n$, which contradicts \eqref{volume-exec}.\\

In general, if two adjacent pillars $n$ and $n+1$ satisfy the PVol constraint then the previous argument shows that $v_i=0$ for all $i=1,\dots,N$. Therefore, the two algorithms TC and PVol are mutually exclusive. This implies that a classical optimisation scheme of TC with the PVol constraint via Lagrange multipliers is not straightforward, to say the least. We thus have to find another way to obtain a solution of the TC algorithm under the PVol constraint.\

From \eqref{tc-algo-1} we see that given $v_n$, the corresponding $v_{n+1}$ depends on $\sum_{i=1}^n v_i$, i.e. the cumulative execution up to $n$, which implies that curve is in general increasing. Therefore, in order to satisfy the constraint of maximum percentage of volume (PVol), if the total volume to execute is large then the algorithm has to be divided into two patterns:
\begin{enumerate}
  \item As long as the constraint of maximum participation rate (PVol) is not reached, we execute the slices according to the Almgren-Chriss recursive formula. This corresponds to the TC pattern.
  \item As soon as the PVol constraint is attained, the algorithm executes the minimum  between the TC curve and PVol curve.
\end{enumerate}
Loosely speaking, we start with a TC algorithm, but once the slices are \emph{saturated} we switch to a PVol algorithm until the end of the execution. However, it can happen that the algo switches back to TC if the PVol curve is bigger at a further pillar; this situation is exceptional though, save for cases where the volume curve presents sharp peaks or gaps.\

It is worth to mention that adding a PVol constraint to IS is essentially the same as adding the constraint for TC and running the TC algorithm backwards.

\subsection{Computing the optimal stopping time for TC}\label{sec:shoot}

Let us describe in detail all the steps of our TC algorithm under PVol constraint. Let $n_0$ be the starting time, $n_1$ the switching time (i.e. when we change from TC to PVol) and $\alpha$ the number of shares we trade at pillar $n_0$.
\begin{enumerate}
\item According to the historical estimates of the available volume at the close auction, plus the desired participation rate, we define the execution at the pillar $n=103$ (the close auction), denoted $v^\sharp$.

\item We compute the Almgren-Chriss algorithm for the residual shares $v^\flat = v^\ast-v^\sharp$ i.e. the shares to execute in continuous, outside of the close auction. We start with $n_0=1$ and $n_1=102$ and launch the TC recursive argument \eqref{tc-algo-1}. Since the algorithm is completely determined by $v_1=\alpha$, it suffices to find the right $\alpha$ such that the cumulative shares at $n_1=102$ are equal to $v^\flat$.

\item We compare the trading curve of the previous step with the PVol curve. If the PVol constraint is satisfied then we are done. If not, we \emph{saturate} pillar $n=102$ with the PVol constraint and redefine the parameters: $v^\flat$ is now the shares to execute outside both pillars 103 and 102, i.e. $v^\flat=v^\flat-v_{102}$, whilst $n_1$ is set to 101, i.e. $n_1=n_1-1$.

\item Eventually, we will obtain a TC curve starting at $n_0=1$ that switches to PVol at $n_1\le102$, satisfying the constraint. Moreover, the algorithm finds the right $\alpha$ at $n_0=1$ such that the total execution from $n=1$ to $n=103$ is equal to $v^\ast$. Remark that the whole algorithm executes TC between $n_0=1$ to $n_1$, PVol between $n_1$ and $n=102$, and the desired participation at the close auction at $n=103$.

\item In order to find the right starting time $n_0$, we define a minimal trading size for each slice, that we denote $\alpha_{min}$. Let $\alpha_0$ be the minimum of the trading curve we have already found in the previous step. If $\alpha_0<\alpha_{min}$ then we advance one pillar, i.e. $n_0$ passes from 1 to 2, and we recompute the trading curve. In order to continue hitting the target we change $\alpha$ for the cumulative trades of the previous step up to pillar $n=2$. We continue until we find the first pillar $n_0$ such that $\alpha_0\ge\alpha_{min}$; notice that in this case $\alpha$ will be the cumulative trades of the previous step up to the pillar $n_0$.

\end{enumerate}
Therefore, $n_1$ is determined by the PVol constraint whilst $n_0$ is determined by the minimal trading size constraint $\alpha_{min}$. Notice however that  the optimal starting pillar $n_0$ is determined after $n_1$, which implies that $n_0$ depends not only on $\alpha_0$ but also on the rest of the parameters, in particular the PVol curve, the participation rate at the close auction and the market impact parameters.\

Observe that there is a systematic way of computing the stopping pillar for an IS algorithm: it corresponds to the \emph{backwards or symmetrical image} of the starting time we computed for the TC algorithm.

\subsection{Numerical results}

\begin{figure}[htbp]
\begin{center}
  \includegraphics[width=4in]{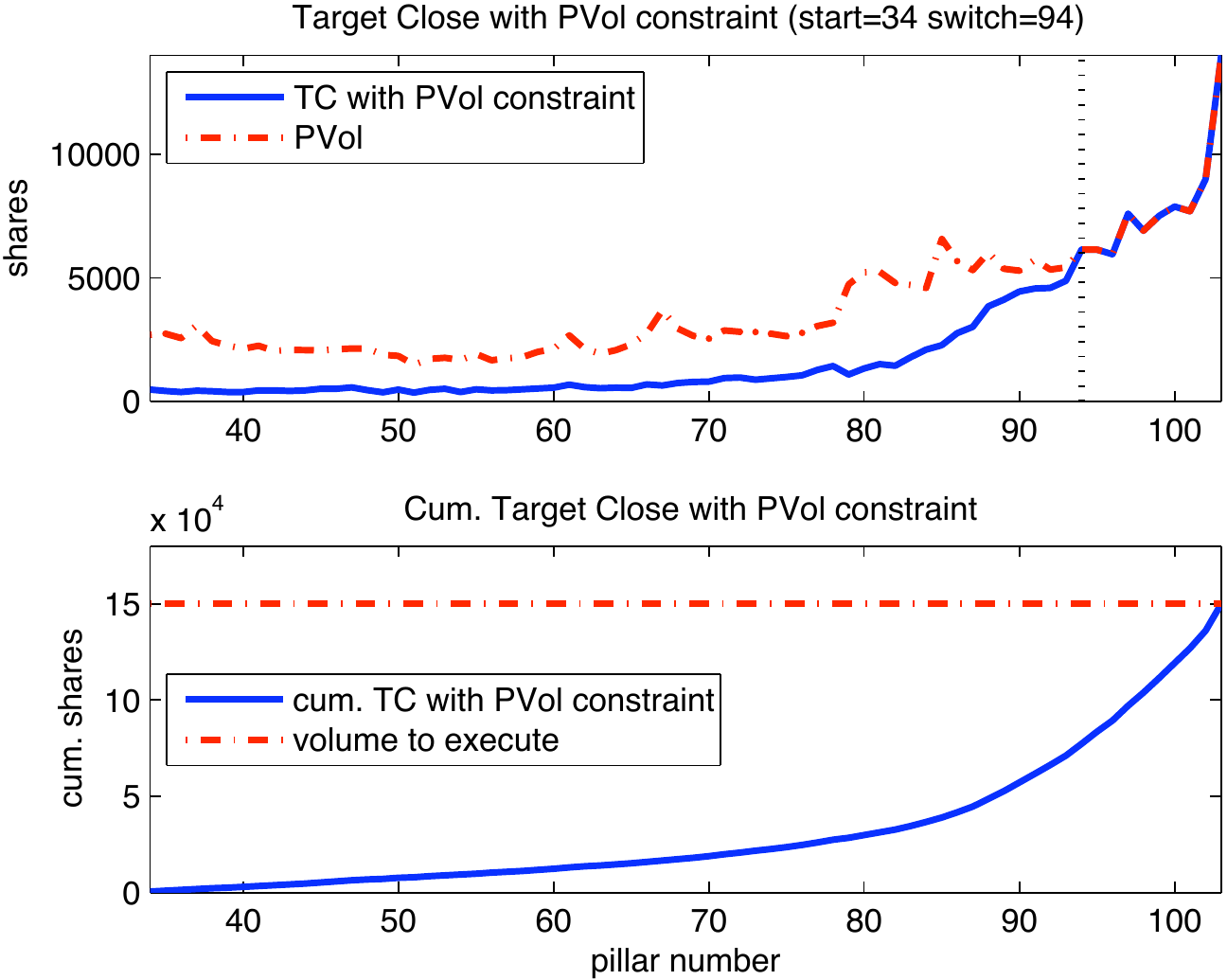}\\
  \caption{\small{Cumulative TC curves under PVol constraint.}}\label{fig-bb-tc-pvol}
\end{center}
\end{figure}

In the first plot of Figure \ref{fig-bb-tc-pvol} we have the TC curve (solid line) under PVol constraint vs the PVol curve (broken line) of stock AIRP.PA (Air Liquide). In the second plot we have the cumulative execution of the TC curve (solid line) under PVol constraint vs the volume to execute $v^\ast$ (broken line). The parameters we used are $v^\ast=150,000$ shares, $\alpha_{min}=500$ shares and a maximum participation of 20\% in both the continuous trading period and the close auction. The historical volatility and volume curves, the market impact parameters $\kappa$ and $\gamma$ and the risk-aversion coefficient $\lambda$ were provided by the Quantitative Research at Cheuvreux - Cr\'edit Agricole.\

In Figure \ref{fig-bb-tc-pvol} we can also observe that the algorithm finds the optimal starting time at pillar $n=34$ (beginning of the horizontal axis), at which it starts to execute the order following the TC algorithm based on the Almgren-Chriss optimisation. At pillar $n=94$ (vertical line) the algorithm switches to PVol in order to satisfy the constraint. Moreover, during the whole execution, the PVol constraint has been satisfied. In the second plot we can see that the TC algorithm under PVol constraint successfully executed the whole order.


\section{Non-Brownian models: self-similar processes}

\subsection{The $p$-variation model}

Let $p>1$ and $y=(y_1,\dots,y_N)\in\R^N$ be a random vector of mean zero. We define the \emph{$p$-variation} of $y$ as
\[
\V_p(y):=\sum_{n=1}^N\E\left[\,\vert \, y_n\vert^p\,\right].
\]
The $p$-variation $\V_p(y)$ and the $l_p$-norm in $\R^N$ are related via
\[
\Vert y\Vert_p=\V_p(y)^{1/p}.
\]
Notice that if $y=(y_1,\dots,y_N)$ is a time series of i.i.d. random variables then the 2-variation reduces to the variance, i.e.
\[
\V_2(y)=\mathrm{Var}(y)\,.
\]
Moreover, it is easy to show that the $p$-variation defines a metric on $\R^N$, and since all norms in $\R^N$ are equivalent there exist $0<\beta_1<\beta_2$ such that
\[
\beta_1\Vert y\Vert_p\le \Vert y\Vert_2\le \beta_2 \Vert y\Vert_p.
\]
Therefore, the variance (i.e. the 2-variation) and the $p$-variation are two equivalent metrics on $\R^N$, and in particular
\begin{equation}\label{2-p-var}
\V_p(y)\sim\V_2(y)^{p/2}\,.
\end{equation}
Now let us define the $p$-variation for a special family of functions of random variables. Let $y=(y_1,\dots,y_N)$ a random vector of mean zero and consider the function $F:\R^N\to \R$ defined as
\[
F(y)=\sum_{n=1}^N y_n.
\]
We define the $p$-variation of $F$ as
\[
\V_p(F)=\sum_{n=1}^N\E\left[\,\vert \, y_n\vert^p\,\right].
\]
Observe that if $y=(y_1,\dots,y_N)$ is a time series of mean zero then $\V_p(F)$ is the sample $p$-th moment of the time series $y$ multiplied by $N$. Finally, for general functions $F$ such that
\[
F-\E(F)=\sum_{n=1}^N y_n
\]
we define their $p$-variation as
\[
\V_p(F):=\V_p(F-\E(F))=\V_p(y).
\]
It is worth o remark that if the random variables $y=(y_1,\dots,y_N)$ are i.i.d. of mean zero and variance 1 then the $2$-variation and the variance of $y$ coincide:
\[
\V_2(F)=\mathrm{Var}(y)\,.
\]

\subsection{Optimal trading algorithms using $p$-variance as risk measure}

Assume that the price dynamics is self-similar, i.e.
\begin{equation}\label{H-discrete}
S_{n+1}=S_n+\sigma_{n+1}\tau^H\varepsilon_{n+1},
\end{equation}
where $H\in(0,1)$ and $(\varepsilon_n)_{1\le n\le N}$ are identically-distributed random variables such that $\E\left[\varepsilon_n\right]=0$, not necessarily independent. We will assume a power market impact of the form
\begin{equation}\label{market-impact-H}
h(v_n)=\kappa\sigma_n\tau^H\left(\frac{v_n}{V_n} \right)^\gamma\,.
\end{equation}

In order to use the $p$-variation as a risk measure, we have to choose the right $p$. From \eqref{H-discrete} we see that if $H=1/2$ we recover the classical Brownian motion, for which the variance is the most common choice for a risk measure. In this case we have $H=1/2$ and $p=2$, which implies that the risk measure is linear in time. This suggests that the correct choice of $p$ is $p=1/H$, since it is the only $p$ that renders the risk $p$-variation as a risk measure linear in time.\

We would like to remark that the idea of a risk measure that is linear in time was also introduced by Gatheral and Schied \cite{citeulike:10363463}, where the risk measure  was the expectation of the time-average. The advantage of our approach is that we do not fix a priori the dynamics of the price process. Indeed, we first find empirically the right exponent of self-similarity $H$ and then we choose the correct risk measure via $p=1/H$.\

In order to derive the recursive formula for a process following \eqref{H-discrete}, we normalise the relative wealth as in the previous case of Brownian motion. Under this framework, the normalised relative wealth of a TC algorithm is
\begin{equation}\label{eq:wealth}
\tilde W =-\sum_{n=1}^{N-1} x_n\sigma_{n+1}\varepsilon_{n+1}+\sum_{n=1}^N \kappa \sigma_n \frac{(x_n-x_{n-1})^{\gamma+1}}{V_n^\gamma}\,.
\end{equation}
We will assume that the process \eqref{H-discrete} is normalised, i.e. $\E\left[\vert\varepsilon_n\vert^p\right]=1$ for all $n$. In the case of Brownian motion ($H=1/2$ and $p=2$) this is equivalent to suppose that the increments $(\varepsilon_n)_{1\le n\le N}$ have variance 1. Under this framework, the mean and $p$-variation of $\tilde W$ are
\[
\mathbb{E}(\tilde W)=\sum_{n=1}^N \kappa\sigma_n \frac{(x_n-x_{n-1})^{\gamma+1}}{V_n^\gamma}\,,\qquad\mathbb{V}_p(\tilde W)=\sum_{n=1}^{N-1} x_n^p\sigma_{n+1}^p\,.
\]
Therefore, the corresponding $p$-functional is
\begin{eqnarray*}
J_p(x_1,\dots,x_N) &=& \mathbb{E}(\tilde W)+\lambda\mathbb{V}_p(\tilde W)\\
 &=& \sum_{n=1}^N \kappa\sigma_n \frac{(x_n-x_{n-1})^{\gamma+1}}{V_n^\gamma}+\sum_{n=1}^N \lambda(n,N)x_n^p\sigma_{n+1}^p\,,
\end{eqnarray*}
where $\lambda(n,N)=\lambda$ if $n<N$ and $\lambda(N,N)=0$. The optimal trading curve is determined by solving
\[
\frac{\partial J_p}{\partial x_n}=0\,,\qquad n=1,\dots,N\,,
\]
i.e.
\[
\kappa\sigma_n(\gamma+1)\frac{(x_n-x_{n-1})^\gamma}{V_n^\gamma}
-\kappa\sigma_{n+1}(\gamma+1)\frac{(x_{n+1}-x_n)^\gamma}{V_{n+1}^\gamma}
+p\lambda(n,N)\sigma_{n+1}^px_n^{p-1}=0\,.
\]
Returning to the variables $v_n$ we get
\[
\kappa\sigma_n(\gamma+1)\left(\frac{v_n}{V_n}\right)^\gamma
-\kappa\sigma_{n+1}(\gamma+1)\left(\frac{v_{n+1}}{V_{n+1}}\right)^\gamma +p\lambda(n,N)\sigma_{n+1}^p\left(\sum_{i=1}^n v_i\right)^{p-1}=0\,.
\]
We thus obtain the recursive nonlinear formula for the optimal TC trading curve for a self-similar process:
\begin{equation}\label{tc-algo-p}
v_{n+1}=V_{n+1}\left[\frac{\sigma_n}{\sigma_{n+1}}\left(\frac{v_n}{V_n}\right)^\gamma+
\frac{p\lambda(n,N)}{\kappa(\gamma+1)}\sigma_{n+1}^{p-1}\left(\sum_{i=1}^n v_i\right)^{p-1}\right]^{1/\gamma}\,.
\end{equation}
If we were interested in the IS algorithm, an argument similar to the previous one would show that the optimal trading curve for IS satisfies
\begin{equation}\label{tc-algo-p-is}
v_{n-1}=V_{n-1}\left[\frac{\sigma_n}{\sigma_{n-1}}\left(\frac{v_n}{V_n}\right)^\gamma+
\frac{p\lambda(n,1)}{\kappa(\gamma+1)}\frac{\sigma_n^p}{\sigma_{n-1}}\left(\sum_{i=n}^N v_i\right)^{p-1}\right]^{1/\gamma}\,,
\end{equation}
where $\lambda(n,1)=\lambda$ if $n>1$ and $\lambda(1,1)=0$.

\subsection{Examples of self-similar processes}

Amongst the class of continuous stochastic processes that admit a discretisation of the form \eqref{H-discrete}, we have three processes in mind: L\'evy processes, fractional Brownian motion and fractal processes (for more details we suggest \cite{citeulike:6335534}, \cite{BOUCH04}, \cite{citeulike:8838678}, \cite{citeulike:2915407} and \cite{citeulike:3417194})).

\begin{itemize}
  \item \textbf{Truncated L\'evy processes}. $p$-stable L\'evy processes are the only self-similar processes satisfying \eqref{H-discrete} with $H=1/p$ and with independent, stationary increments. If $p=2$ we recover the classical Brownian motion. However, for such processes the $p$-th moment is infinite, and as such they cannot be used in our framework. Nevertheless, one can consider the so-called \emph{truncated} L\'evy distributions, which are L\'evy within a bounded interval and exponential on the tails. This allows moments of any order, in particular the $p$-th moment, whilst within the bounded interval we keep the self-similarity given by \eqref{H-discrete}.

  \item \textbf{Fractional Brownian motion}. The fractional Brownian motion is the only self-similar process with stationary, Gaussian increments. Its exponent of self-similarity $H$ is called \emph{Hurst exponent}). If $H=1/2$ we recover the classical Brownian motion. The fractional Brownian motion has moments of all orders, hence the $p$-variation is well-defined and we can apply our model. However, for $H\neq1/2$ the increments are auto-correlated (positively if $H>1/2$ and negatively if $H<1/2$) and our model does not take into account the auto correlations. Therefore, we can consider our model as an approximation when auto correlations are weak with respect to the market impact and the $p$-variance.

  \item \textbf{Multifractal processes}. Multifractal processes are defined as follows. Given a stochastic process $X(t)$ its fluctuation is defined as
      \[
      \delta_lX(t):=X(t+l)-X(t)\,.
      \]
      For any $q>0$ define
      \[
      m(q,l):=\E\left[\vert \delta_lX(t)\vert^q\right]\,.
      \]
      We say that $X(t)$ is multifractal of exponents $\zeta(q)$ if for any $q>0$ there exists $K(q)>0$ such that
      \[
      m(q,l)=K(q)l^{\zeta(q)}\,.
      \]
      In the case where $\zeta(q)$ is linear, i.e. $\zeta(q)=qH$ the process $X(t)$ is called monofractal. If $\zeta(q)$ is not linear then $X(t)$ is called multifractal. Notice that all self-similar processes are monofractal, in particular the fractional Brownian motion and L\'evy processes. However, we will continue to use the term \emph{self-similar}, even for monofractal process, since it is more common in the literature.

\end{itemize}

\subsection{Numerical results}

\begin{figure}[htbp]
\begin{center}
  \includegraphics[width=4in]{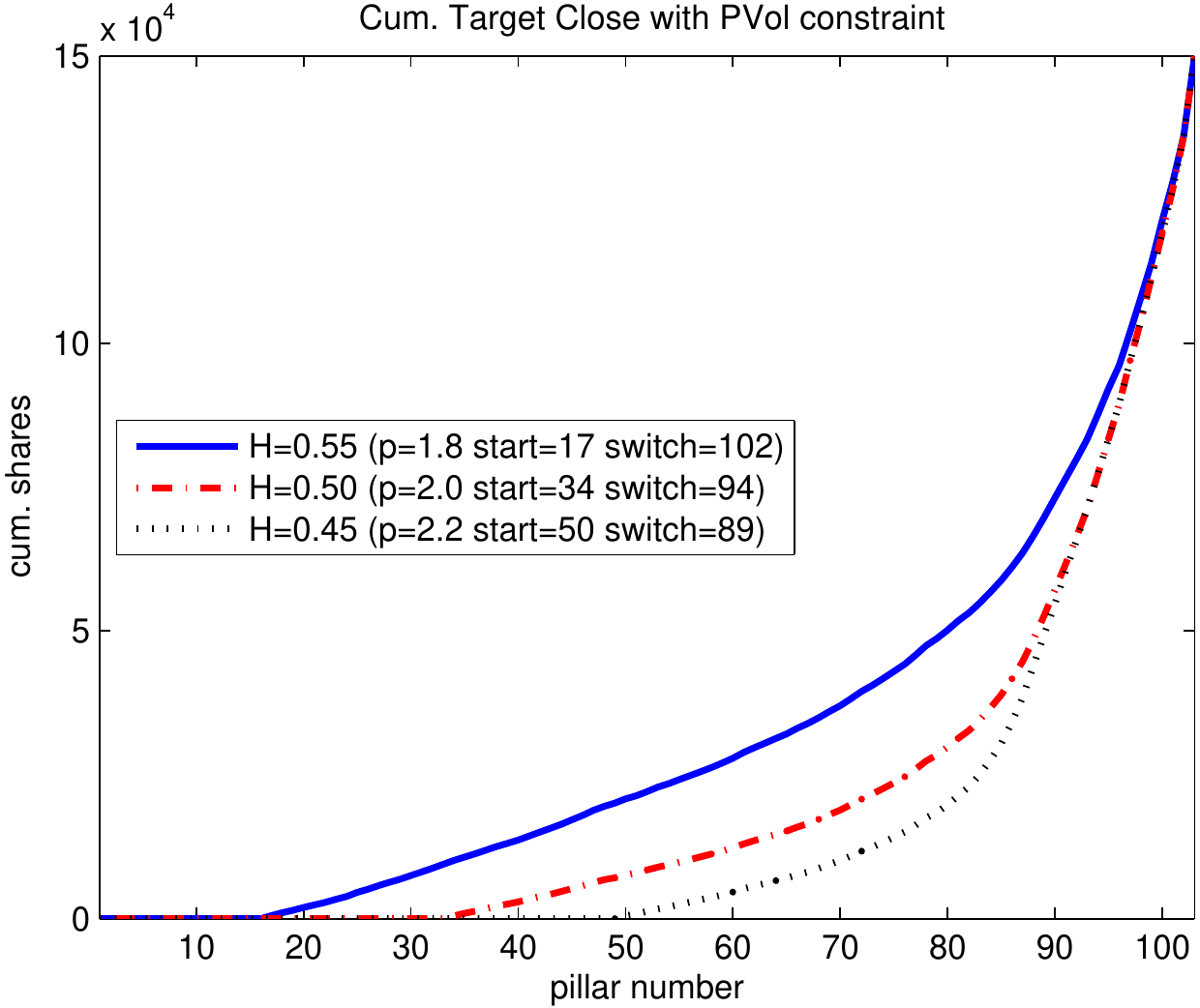}\\
  \caption{\small{Cumulative TC curves under PVol constraint for different values of $H$.}}\label{fig-bb-hurst}
\end{center}
\end{figure}

In Figure \ref{fig-bb-hurst} we plotted three TC curves under the PVol constraint for three different self-similarity exponents $H$, which gives three different $p$'s for the $p$-variation (recall $p=1/H$).

\smallskip

\begin{center}
\begin{tabular}{|c|c|c|c|}
  \hline
  $H$ & $p$ & start pillar & switch pillar \\
  \hline
  0.55 & 1.8 & 17 & 102 \\
  \hline
  0.50 & 2.0 & 34 & 94\\
  \hline
  0.45 & 2.2 & 50 & 89\\
  \hline
\end{tabular}
\end{center}

\smallskip

Our numerical example renders the following evidence, which has been found in all runs we have performed:
\begin{itemize}
  \item If $H$ increases then the starting pillar of the execution decreases, i.e. the execution starts earlier.
  \item If $H$ increases then the pillar at with we switch from TC to PVol increases, i.e. the PVol constraint is saturated later.
\end{itemize}

Since starting the execution later and saturating the PVol constraint earlier is related to higher levels of aggressiveness, we can infer from Figure \ref{fig-bb-hurst} that the level of aggressiveness of TC under the PVol constraint decreases as $H$ increases. This finding is quite natural if we assume that the model is a fractional Brownian motion and $H$ is the Hurst exponent:
\begin{itemize}
  \item For $H<1/2$ the process has negative auto correlations, i.e. it behaves as a mean-reverting process. Therefore, the market impact is reduced because prices go back to their level after an execution. In consequence, we can execute the order faster than in the case of a classical Brownian motion: we start the execution later and we go as fast as possible, and as such we saturate the constraint earlier.
  \item For $H>1/2$ the process has positive auto correlations, i.e. it has a trend. Therefore, the market impact is of paramount importance because if we execute too fast then prices will move in the wrong direction. In consequence, we start the execution earlier and we go as slow as possible, and as such we saturate the constraint later.
\end{itemize}
 In the next section we will study in detail, in the TC algorithm without PVol constraint, the relation between the risk measures of $p$-variation type and both the starting time and the slope at the last pillar.


\section{Assessing the effects of the risk measure}

\begin{figure}[htbp]
\begin{center}
  \includegraphics[width=4in]{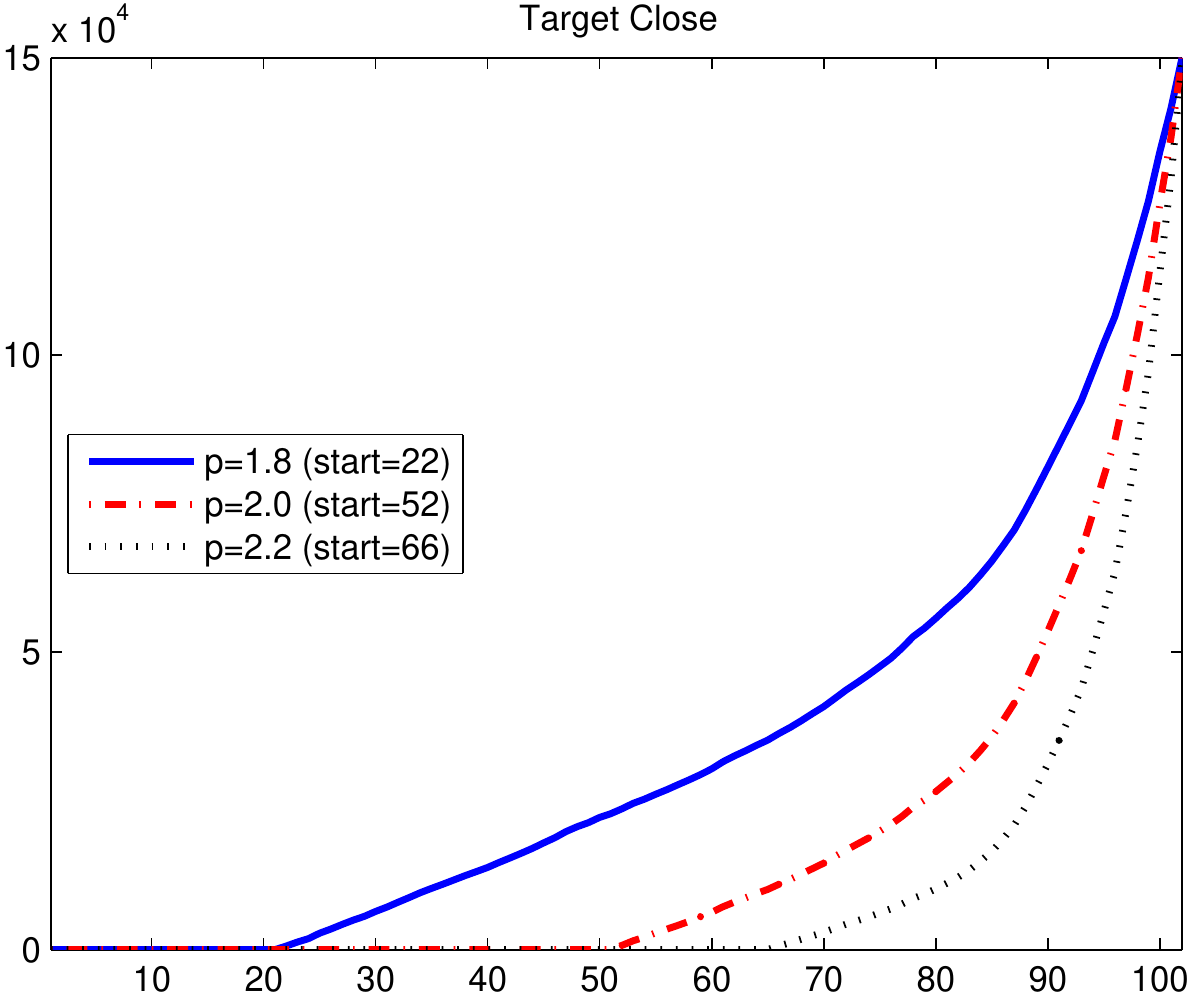}\\
  \caption{\small{Cumulative TC curves without PVol constraint.}}\label{fig-bb-tc-only-cum}
\end{center}
\end{figure}

In this section we consider the TC algorithm without PVol constraint and without participation at the close auction. As expected, the algorithm starts closer to the close without PVol constraint (Figure \ref{fig-bb-tc-only-cum}) than under the restriction (Figure \ref{fig-bb-hurst}). 


We will show that the choice of the parameter $p$ for the $p$-variation plays a crucial role in both the model of the asset we are trading and the weight we give to the market risk with respect to the market impact.

\subsection{Equivalence between risk measures and models}\label{sec:measures}

If we compare the recursive formula \eqref{tc-algo-1} for a Brownian motion with the recursive formula \eqref{tc-algo-p} for a self-similar process, we see that the former can be recovered from the latter by setting  $p=2$ . Therefore, formula \eqref{tc-algo-p} can be derived for a Brownian model when the risk measure is not the variance but the $p$-variation.\

In consequence, \eqref{tc-algo-p} is independent of the model we choose for the asset: assuming a self-similar process, estimating empirically its exponent of self-similarity $H$ and defining $p=1/H$ is equivalent to assuming a Brownian model, choosing $p$ and using the $p$-variation as the risk measure instead of the variance.\

A direct consequence of this analysis is that the risk measure is of paramount importance. Indeed, it not only determines the weight we impose to the market risk but it is implicitly related to a model choice. Indeed, choosing $p>2$ is equivalent to choosing a self-similar process with $H<1/2$, which for the fractional Brownian motion would mean a process with negative auto correlations, i.e. a mean-reverting process. On the other hand, choosing $p<2$ implies $H>1/2$, hence the corresponding fractional Brownian motion has positive auto correlations, i.e. it has a trend.\\

In summary:
\begin{proposition}\label{prop-risk-models}

In order to obtain the recursive formula \eqref{tc-algo-p} for a TC algorithm via an Almgren-Chriss optimisation, the following two paths are equivalent:
\begin{enumerate}
\item Assuming a self-similar model for the asset, calibrating empirically its exponent of self-similarity $H$ and choosing the $p$-variation as the risk measure with $p=1/H$.
\item Assuming a Brownian motion model for the asset and choosing the $p$-variation as the risk measure.
\end{enumerate}

\end{proposition}

\subsection{Risk measures, starting times and slopes for the TC algorithm}

The aggressiveness of a TC algorithm can be measured in terms of both the starting time and the slope of the trading curve: an algorithm is more aggressive if it starts later and it executes \emph{faster} the trades i.e. it puts more shares per pillar and the rate of change between consecutive pillars is bigger.\\

\begin{figure}[htbp]
\begin{center}
  \includegraphics[width=4in]{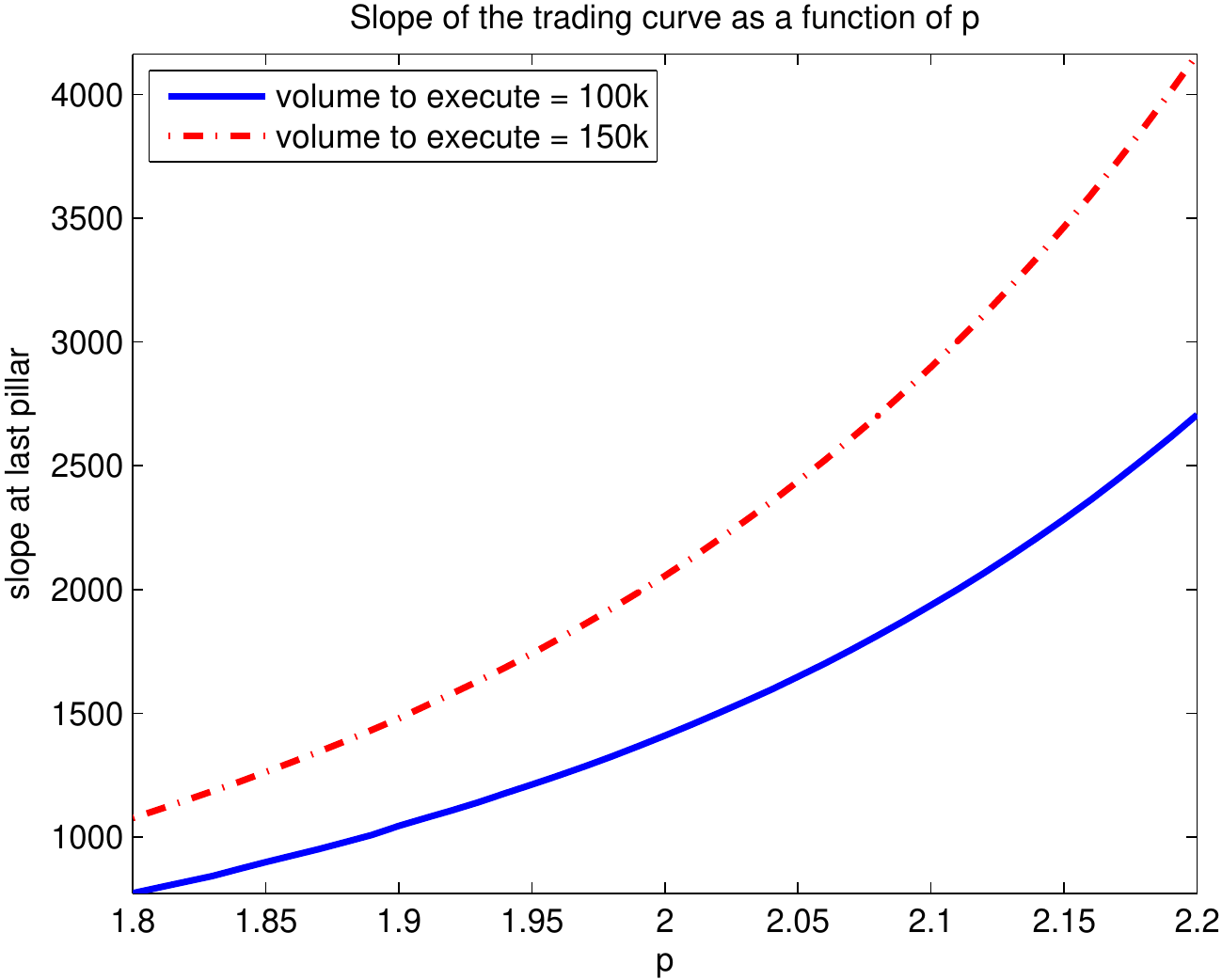}\\
  \caption{\small{The slope at the last pillar is increasing in $p$.}}\label{fig-bb-tc-slope}
\end{center}
\end{figure}

 Our numerical simulations confirm the analytical results of the previous section: the optimal starting time and the slope of the trading curve are both monotone increasing in $p$ (see Figures \ref{fig-bb-start-time} and \ref{fig-bb-tc-slope}, respectively). Under this framework, $p$ can be viewed as a tuning parameter for aggressiveness:

\begin{figure}[htbp]
\begin{center}
  \includegraphics[width=3.8in]{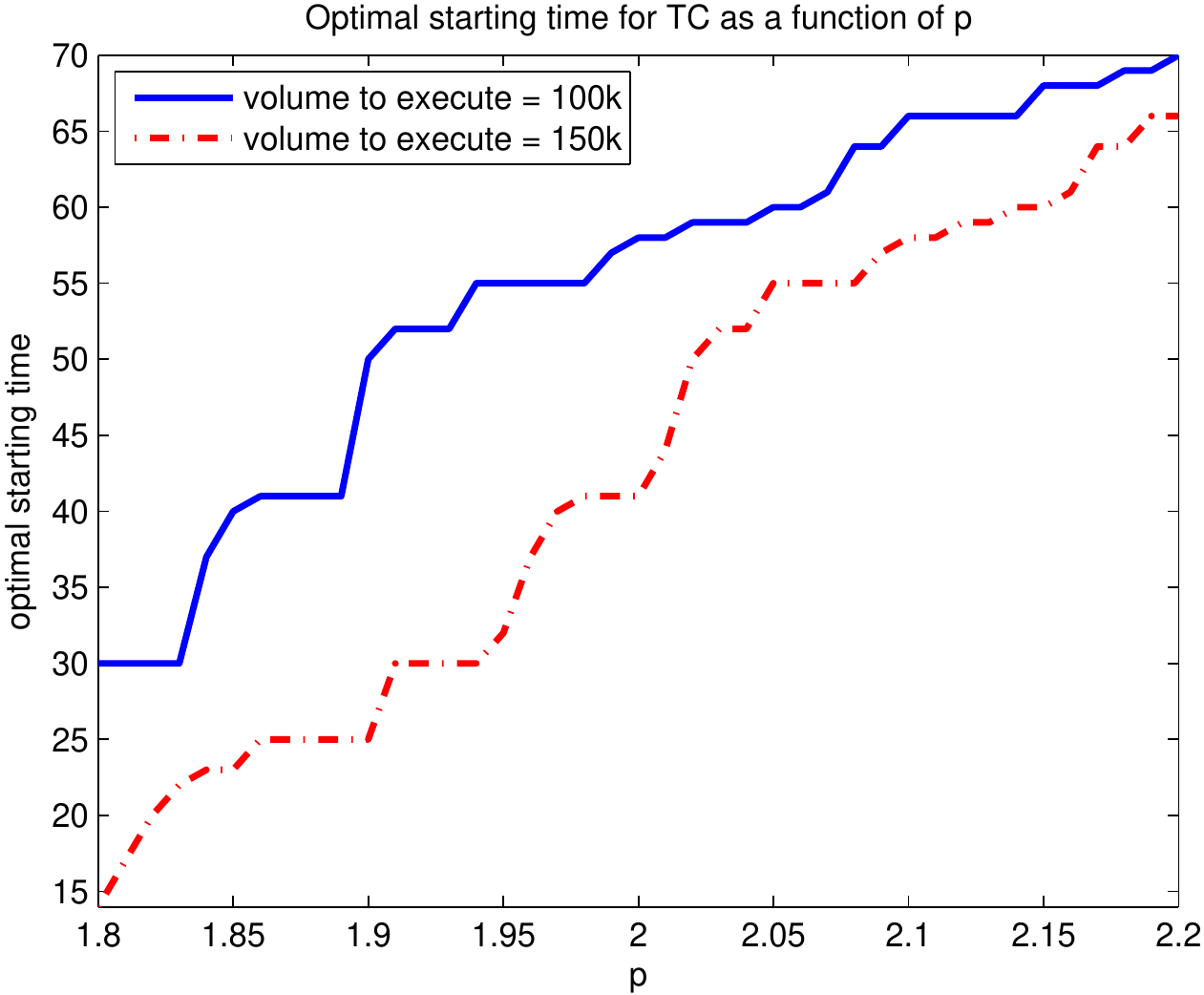}\\
  \caption{\small{The optimal start time is increasing in $p$.}}\label{fig-bb-start-time}
\end{center}
\end{figure}

\begin{proposition}\label{prop-p-aggressive}

The parameter $p$ in the recursive formula \eqref{tc-algo-p} measures the level of aggressiveness of the TC algorithm. More precisely:
\begin{enumerate}
\item $p$ increases $\Longleftrightarrow$ the optimal stating time increases.
\item $p$ increases $\Longleftrightarrow$ the slope of the cumulative trading curve at the last pillar increases.
\end{enumerate}
In consequence, $p$ increases if and only the TC algorithm is more aggressive, i.e. it starts the execution later and executes more at each pillar.
\end{proposition}

\subsection{Implied $p$-variation for CAC40 and link with liquidity}\label{sec:implied}

\begin{figure}[htbp]
\begin{center}
  \includegraphics[width=3.7in]{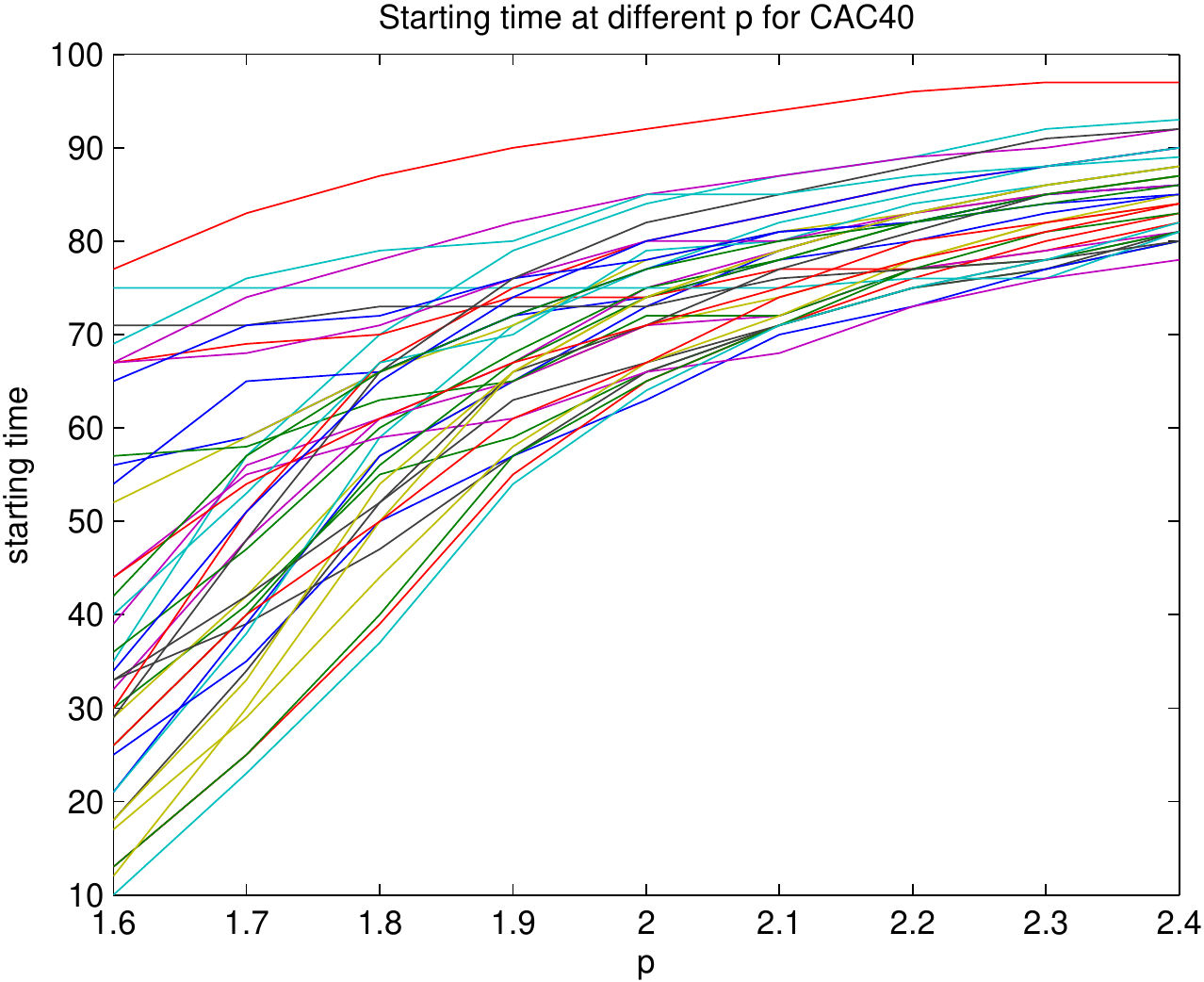}\\
  \caption{\small{Cumulative TC curves.}}\label{fig-bb-cac40_40}
\end{center}
\end{figure}

In Figure \ref{fig-bb-cac40_40} we plotted the starting times as a function of $p$ for 39 out of 40 stocks in the CAC40 index.\\

The $p$ parameter has been considered so far as an input whilst the starting time $n_0$ was an output. However, since $p\mapsto n_0(p)$ is increasing, we can consider the inverse problem: given a starting time $n^\sharp$ there is a $p$ such that the trading curve for the TC algorithm executes the total number of shares between $n_0$ and the last trade before the closing auction. The $p$ is not unique because $n_0$ can only take discrete values, which implies that $p\mapsto n_0(p)$ is piecewise constant. However, it can be rendered unique if we define the implied $p$ as
\[
p :=\sup\{p'\,:\, n_0(p')=n^\sharp\}\,.
\]\\

In consequence, we have an implied $p$ for the CAC40 index: given a common starting time $n^\sharp$, for each stock in the CAC40 index we find their $p$ such that $n_0(p)=n^\sharp$. For the numerical simulations in Figure \ref{fig-bb-cac40_40} we chose $n^\sharp=77$, which corresponds to the opening of the NYSE in the US, a very important time for European traders. We supposed that, for each name on the CAC40 index, we have to execute 6\% of the total daily volume. For volume curves, volatility curves and market impact parameters we used those provided by the Quantitative Research at Cheuvreux.\\

The statistics of the implied $p$ are summarised in the next table:

\smallskip

\begin{center}
\begin{tabular}{|c|c|c|c|}
\hline
\textbf{minimum} & 1.60 & \textbf{quantile 25\%} & 2.00\\
\hline
\textbf{maximum} & 2.40 & \textbf{median} & 2.10 \\
\hline
\textbf{mean} & 2.11& \textbf{quantile 75\%} & 2.20\\
\hline
\textbf{std deviation} & 0.17 &  & \\
\hline
\end{tabular}
\end{center}

\small

The implied $p$ can be very useful for executing portfolios: we can synchronise all assets in our basket, so that all executions start at the same time $n^\sharp$. In this way we can assess and compare the market impact of the individual executions on the same ground, just as options traders use the implied volatility for that purpose. Of course, for real portfolio execution this is far from optimal, but at least it is a first step towards a systematic, quantitative measure of portfolio execution.\\

\begin{figure}[htbp]
\begin{center}
  \includegraphics[width=3.9in]{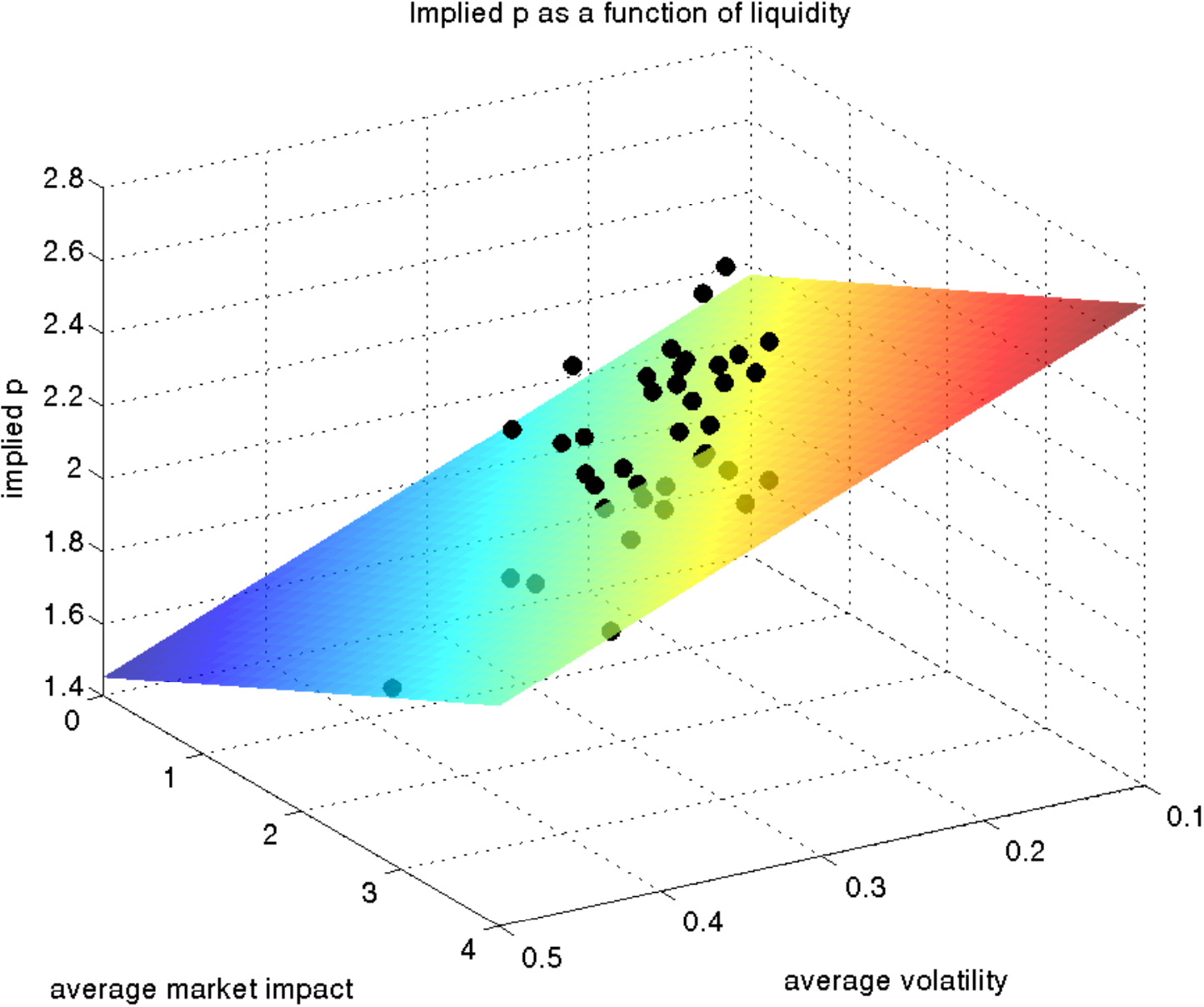}\\
  \caption{\small{Implied $p$ and liquidity.}}\label{fig-bb-liq-surf-color}
\end{center}
\end{figure}

Finally, the implied $p$ can be viewed as a measure of the joint impact of the volatility and the liquidity, the latter modelled as the market impact. In order to illustrate that fact, we performed a linear regression on the implied $p$ with respect to the average volatility per year and the average market impact per pillar given by \eqref{market-impact-H}. The coefficients of the linear regression are:
\begin{equation}\label{eq-regression}
\textnormal{implied $p$} \quad\cong\quad 2.35 \quad+\quad 0.14\times\textnormal{market impact}
\quad-\quad1.79\times\textnormal{volatility}
\end{equation}
with $R^2=0.27$.  The results of \eqref{eq-regression} and Figure \ref{fig-bb-liq-surf-color} can be interpreted as follows:
\begin{itemize}
\item If the market impact decreases then the TC algorithm will start the execution later because the execution will have a smaller effect on moving the price in the wrong direction. However, since the starting pillar has been fixed, the TC algorithm compensates this fact by playing less aggressively, i.e. by decreasing $p$.
\item If the volatility increases then the TC algorithm would like to start later in order to avoid paying the market risk. However, since the starting time is fixed, the algorithm plays less aggressive and thus it decreases $p$.
\end{itemize}

\section{Final remarks}

\paragraph{What is the role of the Hurst exponent in the model?}\

As it was stated in Section \ref{sec:measures}, the choice of $p$ in the $p$-variation can come from two sources, either computing the Hurst exponent $H$ of the process and choosing $p= 1/H$, either by assuming a Brownian motion and choosing the value of $p$ that fits the trader's schedule.\

In practice, the Hurst exponent is not very robust statistically, which means that the right $p$ is to be implied, not computed directly from $H$. One solution is to choose the starting time $n_0$ of the TC algorithm and find the $p$ that forces the optimal curve to start at $n_0$, just as we did in Section \ref{sec:implied}. Another solution is to use $p$ as a fine-tuning parameter: since the optimal curves are defined ex-ante, changing $p$ dynamically during the execution gives room to capitalise potential trends. For example, if the trader chose $p=2$ then they assumed that the prices would behave as a martingale. If a mean-reversion dynamic in the prices is observed the trader can change for a new parameter $p_1 > 2$, whilst if a trend-following dynamic is spotted the new parameter would be $p_2 < 2$. The exact values of $p_1$ and $p_2$ depend on the risk budget of the trader, but in both cases there is a clear interpretation in terms of the Hurst exponent $H$: the process is mean-reverting for $H<1/2$, a martingale for $H=1/2$ and has a trend for $H>1/2$.\

\paragraph{Why $p$-variation instead of the $p$-th moment of the wealth?}\

If we choose the central $p$-th moment as a risk measure, even in the case of i.i.d. random variables $(\varepsilon_n)_{1\le n\le N}$ we would have several non-zero cross terms of the form
\[
x_i^a x_j^b\mathbb{E}[\varepsilon_i^a\varepsilon_j^b]\,,\quad i\neq j\,,\quad a+b=p\,.
\]
If $p$ is an integer then the wealth $W$ still has an explicit expression, which would lead to an explicit cost functional $J_p$ and therefore explicit equations for the optimal trading curve on the variables $(v_0,\dots,v_{N+1})$ of the form
\[
v_n = f(v_0,v_1,\dots,v_n,\dots,v_N,v_{N+1})\quad\textnormal{for all $n$.}
\]
These equations can be solved computationally, although the numerical algorithm is no longer explicit and recursive but implicit. If $p$ is not an integer then the wealth is no longer explicit, which means that the computational effort needed to find the optimal trading curve is bigger.\
 
Choosing the $p$-variation instead of the $p$-th moment present several advantages. First, the algorithm for the optimal trading curve is explicit and recursive of the form
\[
v_n = f(v_0,v_1,\dots,v_{n-1})\quad\textnormal{for all $n$.}
\]
This explicitness allows to study and interpret the effect of the parameters on the shape of the trading curves. Second, the $p$-variation can be interpreted as the $\ell^p$-norm of a vector intimately related to the wealth $W$. Indeed, if we define the vectors
\begin{eqnarray*}
X &:=& (x_1\sigma_2,\dots,x_{N-1}\sigma_N)\,,\\
E &:=& (\varepsilon_2,\dots,\varepsilon_N)
\end{eqnarray*}
and consider the relative wealth $\tilde W$ of the TC algorithm given in \eqref{eq:wealth}, we obtain the ``vectorial'' identities
\[
\tilde W - \mathbb{E}(\tilde W) = X\cdot E\quad\textnormal{and}\quad\mathbb{V}_p(\tilde W)^{1/p} = \Vert X \Vert_p\,.
\]
Third, the $p$-variation lets us to model the price as a general self-similar process (e.g. fractional Brownian Motion and L\'evy process) without much of a fuss. Indeed, there is only a single parameter to adjust, namely $p$, which can be found as the inverse of the Hurst exponent $H$, i.e. $p=1/H$. If we use the $p$-th moment instead then the equations become much more involved due to the autocorrelations. Fourth, the $p$-variation can be seen as an approximation of the the $p$-th moment, in which the cross terms are neglected.


\paragraph{Why adding a new risk parameter $p$ when we already have $\lambda$?}\

There are two main reasons. The first is that, in the recursive formulas \eqref{tc-algo-p} and \eqref{tc-algo-p-is} for the TC algorithm, $\lambda$ enters linearly whilst $p$ enters as a power. This implies that the effect on the optimal execution curve is stronger varying $p$ than $\lambda$, in particular the acceleration near the closing time. In consequence, playing with $\lambda$ or $p$ is similar but not equivalent, since the nonlinear behaviour of $p$ cannot be replicated using the linear parameter $\lambda$. The second reason is that $\lambda$ enters into play as a Langrange multiplier in the mean-variance optimisation, and as such it is not evident to choose a standard value of this risk adversion parameter. On the contrary, $p$ has a neat interpretation via the Hurst exponent using the identity $p=1/H$. Moreover, there is a standard value for $p$: if we assume martingality on the price process then $p=2$.

\section*{Acknowledgements}

Most of this research was done when both authors were working in the Department of Quantitative Research at Cr\'edit Agricole Cheuvreux (now Kepler Cheuvreux); they are thankful for the help and support provided by the firm and the team. Mauricio Labadie would like to thank EXQIM as well for the help and support during the final stages of the paper. The authors would also like to thank an anonymous referee for their remarks and suggestions, which helped to improve this paper.


\bibliographystyle{apalike}
\bibliography{lehalle}

\begin{thebibliography}{}

\bibitem[Abergel et~al., 2012]{citeulike:10363473}
Abergel, F., Bouchaud, J.-P., Foucault, T., Lehalle, C., and Rosenbaum, M.,
  editors (2012).
\newblock {\em {Market Microstructure Confronting Many Viewpoints}}.
\newblock Wiley.

\bibitem[Almgren, 2003]{almgren03}
Almgren, R.~F. (2003).
\newblock {Optimal execution with nonlinear impact functions and
  trading-enhanced risk}.
\newblock {\em Applied Mathematical Finance}, 10(1):1--18.

\bibitem[Almgren and Chriss, 2000]{OPTEXECAC00}
Almgren, R.~F. and Chriss, N. (2000).
\newblock {Optimal execution of portfolio transactions}.
\newblock {\em Journal of Risk}, 3(2):5--39.

\bibitem[Bacry et~al., 2001]{citeulike:6335534}
Bacry, E., Delour, J., and Muzy, J.~F. (2001).
\newblock {Multifractal random walk}.
\newblock {\em Physical Review E}, 64(2):026103+.

\bibitem[Bertsimas and Lo, 1998]{BLA98}
Bertsimas, D. and Lo, A.~W. (1998).
\newblock {Optimal control of execution costs}.
\newblock {\em Journal of Financial Markets}, 1(1):1--50.

\bibitem[Bouchard et~al., 2011]{citeulike:5797837}
Bouchard, B., Dang, N.-M., and Lehalle, C.-A. (2011).
\newblock {Optimal control of trading algorithms: a general impulse control
  approach}.
\newblock {\em SIAM J. Financial Mathematics}, 2:404--438.

\bibitem[Bouchaud, 2010]{bouchaud10}
Bouchaud, J.-P. (2010).
\newblock {Price Impact}.

\bibitem[Bouchaud and Potters, 2004]{BOUCH04}
Bouchaud, J.~P. and Potters, M. (2004).
\newblock {\em {Theory of Financial Risk and Derivative Pricing: From
  Statistical Physics to Risk Management}}.
\newblock Cambridge University Press.

\bibitem[Cont et~al., 1997]{citeulike:10665597}
Cont, R., Potters, M., and Bouchaud, J.-P. (1997).
\newblock {Scaling in Stock Market Data: Stable Laws and Beyond}.
\newblock {\em Social Science Research Network Working Paper Series}.

\bibitem[Embrechts, 2002]{citeulike:8838678}
Embrechts, P. (2002).
\newblock {\em {Selfsimilar Processes (Princeton Series in Applied
  Mathematics)}}.
\newblock Princeton University Press.

\bibitem[Gatheral and Schied, 2012]{citeulike:10363463}
Gatheral, J. and Schied, A. (2012).
\newblock {Dynamical models of market impact and algorithms for order
  execution}.
\newblock In Fouque, J.-P. and Langsam, J., editors, {\em Handbook on Systemic
  Risk (Forthcoming)}. Cambridge University Press.

\bibitem[Gatheral et~al., 2010]{citeulike:6699563}
Gatheral, J., Schied, A., and Slynko, A. (2010).
\newblock {Transient Linear Price Impact and Fredholm Integral Equations}.
\newblock {\em Social Science Research Network Working Paper Series}.

\bibitem[Gu\'{e}ant et~al., 2011]{citeulike:9272221}
Gu\'{e}ant, O., Lehalle, C.-A., and Fernandez-Tapia, J. (2011).
\newblock {Dealing with the inventory risk}.
\newblock Technical report.

\bibitem[Lehalle, 2009]{citeulike:5094012}
Lehalle, C.-A. (2009).
\newblock {Rigorous Strategic Trading: Balanced Portfolio and Mean-Reversion}.
\newblock {\em The Journal of Trading}, 4(3):40--46.

\bibitem[Lehalle, 2012]{citeulike:10363469}
Lehalle, C.-A. (2012).
\newblock {\em {Market Microstructure knowledge needed to control an intra-day
  trading process}}.
\newblock Cambridge University Press.

\bibitem[Mandelbrot and Hudson, 2004]{citeulike:2915407}
Mandelbrot, B. and Hudson, R.~L. (2004).
\newblock {\em {The (Mis)behavior of Markets}}.
\newblock Basic Books, first printing edition.

\bibitem[Mantegna and Stanley, 1994]{citeulike:3417194}
Mantegna, R.~N. and Stanley, H.~E. (1994).
\newblock {Stochastic Process with Ultraslow Convergence to a Gaussian: The
  Truncated L\'evy Flight}.
\newblock {\em Physical Review Letters}, 73(22):2946--2949.

\bibitem[M\"{u}ller et~al., 1990]{citeulike:8524050}
M\"{u}ller, U.~A., Dacorogna, M.~M., Olsen, R.~B., Pictet, O.~V., Schwarz, M.,
  and Morgenegg, C. (1990).
\newblock {Statistical study of foreign exchange rates, empirical evidence of a
  price change scaling law, and intraday analysis}.
\newblock {\em Journal of Banking \& Finance}, 14(6):1189--1208.

\bibitem[Pag\`{e}s et~al., 2012]{citeulike:5177512}
Pag\`{e}s, G., Laruelle, S., and Lehalle, C.-A. (2012).
\newblock {Optimal split of orders across liquidity pools: a stochatic
  algorithm approach}.
\newblock {\em SIAM Journal on Financial Mathematics (Forthcoming)}.

\bibitem[Perko, 2001]{book:perko}
Perko, L. (2001).
\newblock {\em {Differential Equations and Dynamical Systems}}.
\newblock Springer Verlag.

\bibitem[Stoer and Bulirsch, 1983]{book:stoer}
Stoer, J. and Bulirsch, R. (1983).
\newblock {\em {Introduction to numerical analysis}}.
\newblock Springer Verlag.

\bibitem[Xu and Gen\c{c}ay, 2003]{citeulike:10665572}
Xu, Z. and Gen\c{c}ay, R. (2003).
\newblock {Scaling, self-similarity and multifractality in FX markets}.
\newblock {\em Physica A: Statistical Mechanics and its Applications},
  323:578--590.

\end{thebibliography}

\end{document}